\documentclass[12pt,preprint]{aastex}

\usepackage{natbib}

\newcommand {\BD}{BD+30$^{\circ}$36939}

\shorttitle{NLTE model of NGC6543's central star}
\shortauthors{Georgiev et al.}

\begin{document}


\title{NLTE Model of NGC 6543's Central Star and its Relation with the 
Surrounding
Planetary Nebula}


\author{Georgiev  L. N.}
\affil{Instituto de Astronom\'{\i}a, Universidad Nacional Aut\'onoma de 
M\'exico, CD. Universitaria,
Apartado Postal 70-264, 04510 M\'exico DF, M\'exico}
\email{georgiev@astroscu.unam.mx}

\and
\author{Peimbert M.}
\affil{Instituto de Astronom\'{\i}a, Universidad Nacional Aut\'onoma de 
M\'exico, CD. Universitaria,
Apartado Postal 70-264, 04510 M\'exico DF, M\'exico}

\and
\author{Hillier D. J. }
\affil{Physics and Astronomy Department, University of Pittsburgh, 3941 O'Hara 
Street, Pittsburgh PA, USA}

\and 
\author{Richer M. G. }
\affil{Instituto de Astronom\'{\i}a, Universidad Nacional Aut\'onoma de 
M\'exico, CD. Universitaria,
Apartado Postal 70-264, 04510 M\'exico DF, M\'exico}

\author{Arrieta A. }
\affil{Universidad Iberoamericana, Departamento de F\'\i sica y
Matem\'aticas, Av. Prolongaci\'on Paseo de la Reforma 880, Lomas de
Santa Fe, CP 01210, M\'exico, D.F., M\'exico}

\and

\author{Peimbert A.}
\affil{Instituto de Astronom\'{\i}a, Universidad Nacional Aut\'onoma de 
M\'exico, CD. Universitaria,
Apartado Postal 70-264, 04510 M\'exico DF, M\'exico}

\begin{abstract}
We analyze the chemical composition of the central star of the planetary 
nebula NGC 6543 based upon a detailed NLTE model of its stellar wind. The logarithmic 
abundances by number are H=12.00, He=11.00, C=9.03, N=8.36, O=9.02, Si=8.19, P=5.53, 
S=7.57 and Fe=7.24. Compared with the solar abundances, most of the elements 
have solar composition with respect to hydrogen except C which is overabundant by 0.28 
dex and Fe which is depleted by $\sim 0.2$\,dex. Contrary to most previous work, we find 
that the star is not H-poor and has a normal He composition.  These abundances are compared with those found in the diffuse X-ray plasma and the nebular gas.  Compared to the plasma emitting in diffuse X-rays, the stellar wind is much less depleted in iron.  Since the iron depletions in the nebular gas and X-ray plasma are similar, we conclude that the plasma emitting diffuse X-rays is derived from the nebular gas rather than the stellar wind.  
Excellent agreement is obtained between the abundances in the stellar wind and the nebular recombination line 
abundances for He, C, and O relative to H. On the other hand, the derived stellar N 
abundance is smaller than the nebular N abundance derived from recombination lines and agrees with the abundance found from collisionally-excited lines.
The mean temperature variation determined by five different methods indicates that 
the difference in the nebular abundances between the recombination lines and 
collisionally excited lines can be explained as due to the temperature variations 
in a chemically homogeneous medium. 
\end{abstract}


\keywords{stellar winds: planetary nebulae: individual(\objectname{NGC 6543}}

\section{Introduction}
	The generally accepted scenario of planetary nebula formation is the 
interaction 
between the stellar material expelled at low velocity during the AGB phase (slow 
wind) and 
at high velocity at some later moment \citep[fast wind;][]{kwoketal1978}. A direct consequence of this 
scenario 
is the presence of shock fronts propagating inward into the fast wind and outward 
into the slow wind. The 
shocked gas of the fast wind should be heated to temperatures that correspond 
to its kinetic energy. 
 The typical fast wind velocity is $\sim 1000$ km/s and the mass loss rate is 
$1\times 10^{-8}$ M$_\odot$/yr,
which imply temperatures $\sim 10^7$ K. At these temperatures the gas should 
emit in X-rays and 
that emission has been detected recently \citep{2001ApJ...553L..69C,2005AIPC..804..157G}. 
However, the observed temperatures of the X-ray gas are $1 - 3 \times 10^6$ K, 
an order of magnitude lower than expected. Several scenarios have been proposed 
to explain this
discrepancy \citep{2003ApJ...583..368S}, but none of them are easily 
accepted as the 
best one. 

In our previous paper \citep{2006ApJ...639..185G}, we tried to learn more of the X-ray 
emitting region 
using optical coronal lines of iron. Unfortunately, these lines were not detected, 
which set an upper 
limit to the
iron composition of the X-ray emitting gas, a limit far below the solar 
metallicity. This result 
was confirmed by \citet{2006IAUS..234..169K} who did not detect iron lines in 
their high resolution 
X-ray spectrum of \BD. The peculiar chemical composition of the X-ray emitting 
gas leads
to a possible trace of its origin. If the compositions of the stellar wind, 
the nebula, and the X-ray emitting gas are compared, one can decide which of 
them are related,
and that might indicate the origin of the X-ray gas and perhaps provide clues to the physical processes leading to its low 
temperature. 

Some PNe, NGC 6543 among them, have iron abundances which are very depleted with respect to the solar value \citep{1999A&A...347..967P,2007ApJ...654.1068M}, most probably due to depletion into dust grains. The depleted iron content in both the nebular and the X-ray emitting gas points to a possible relation between them. Unfortunately, during the AGB phase of the evolution, iron atoms can capture slow neutrons (the s-processes) \citet{1999ARA&A..37..239B}. The s-processes depletes iron and one needs a good estimate of the composition of the stellar wind before arriving at any conclusion on the relation between the hot gas and its surroundings.

In this paper, we model the stellar wind of the central star of NGC~6543, one of the brightest planetary nebulae emitting X-rays, to derive its chemical composition.  Section~2 presents the observed spectrum and Section~3 describes the selection of the model parameters.  In Sections 4 and 5, we compare the chemical composition of the stellar wind with that of the nebular gas and the X-ray plasma, respectively.  Section 6 considers the evolution of the progenitor of NGC 6543 while Section 7 presents our conclusions.

\section{Observations}
To construct a good stellar atmosphere model that includes the stellar wind, we need information from as wide a range of wavelengths as possible.  We constructed the spectrum of NGC 6543's central star from four sources, three covering the ultraviolet and one covering the optical.

First, we extracted the FUSE spectrum (R $\sim$ 15000 - 20000) by program Q108 (Vidal-Madjar, A) obtained on 2000 October 1 at 21h 51min.  The spectrum was constructed from SiC 2A, SiC 2B, LiF 1A and LiF 1B fragments.  All parts of the spectrum were 
shifted in wavelength and scaled to match the LiF 1B segment. The resulting spectrum was wavelength shifted and scaled to match the overlapping part of the STIS spectrum.  Second, two HST STIS spectra (R $\sim$ 100000) from program 9736 (P.I. R. Williams), taken on 2004 April 7, were used to cover the wavelength region from
1190 to 1500\AA. Unfortunately, the third spectrum from the same program, which 
covers the \ion{C}{4} 1548/50 \AA \ region, has a problem with the wavelength 
calibration and is not useful. Third, we used IUE spectrum SWP33504 (R $\sim$ 10000) to cover the region 1500 - 2000 \AA. The two LWP high resolution spectra of NGC 6543 are too noisy and we did not use them in the diagnostic procedure. These three fractions of the spectrum were finally combined, and the resulting spectrum was used as a temperature diagnostic and for determining the reddening. After the temperature and reddening were determined, we normalized the observed spectrum to the predicted model continuum scaled to match the line free region between 1750 \AA \ and 1900 \AA. 

Finally, the fourth set of spectra were in the optical and consisted of spectra 
obtained by us at the Observatorio Astron\'omico Nacional in the Sierra of San Pedro Martir (SPM) using the REOSC echelle spectrograph 
(R $\sim$ 20000).
A blue spectrum was obtained on 2004 May 26. The data and 
their reduction are described in \citet{2006ApJ...639..185G}. 
This spectrum was obtained to maximize
S/N. All of the strong nebular lines are saturated and the spectrograph 
was set so that $H_\alpha$
was outside the observed wavelength interval. Red spectra were obtained on 2004 June 6 with short
exposures so that the hydrogen and oxygen lines were not saturated. The spectrum 
was reduced using MIDAS.  After the standard CCD reductions, the spectrum was extracted only in the part of the slit covered by the star. Two additional spectra were extracted in adjoining windows. The signal from these windows was averaged, scaled, and subtracted from the stellar spectrum. As a result, most of
the nebular spectrum was removed, except in the cores of the strong lines arising from both the nebula and the central star.  For these stellar lines, the profiles are severely perturbed, but, keeping in mind that the stellar lines are much broader than the nebular lines, the residuals of the strong nebular lines do not affect the wings of the stellar lines, and so they may still be compared with model line profiles. This procedure worked well for the blue spectrum. The 
red spectrum is strongly contaminated by scattered light from $\mathrm H_\alpha$. We corrected the spectrum for this contamination, but we expect $\mathrm H_\alpha$ to be over-corrected. Therefore, the observed intensity of $\mathrm H_\alpha$ is expected to be less than the model prediction. After the spectrum was corrected for the nebular contamination, it was normalized to the continuum by fitting polynomials to the line-free regions.

\section{Model}

	The central star of NGC 6543 is known to be variable. A detailed analysis of the spectral 
variability was published by \citet{2007MNRAS.tmp..882P}. They found that ``Mostly the flux changes
are at $\sim$ 10 to 20 per cent of the continuum level and occur over localized
blueward velocity regions, as opposed to the flux increasing
or decreasing simultaneously over the entire absorption trough.'' For the rest of the spectrum
the authors concluded that ``Any changes in the other spectral lines are rather subtle if present at all.'' 
Therefore, our
analysis, which is based upon the overall line profile and, in most cases, on 
weak absorption lines, is not affected by the variability.

\subsection{Helium abundance}

\begin{figure}
\epsscale{.80}
\plotone{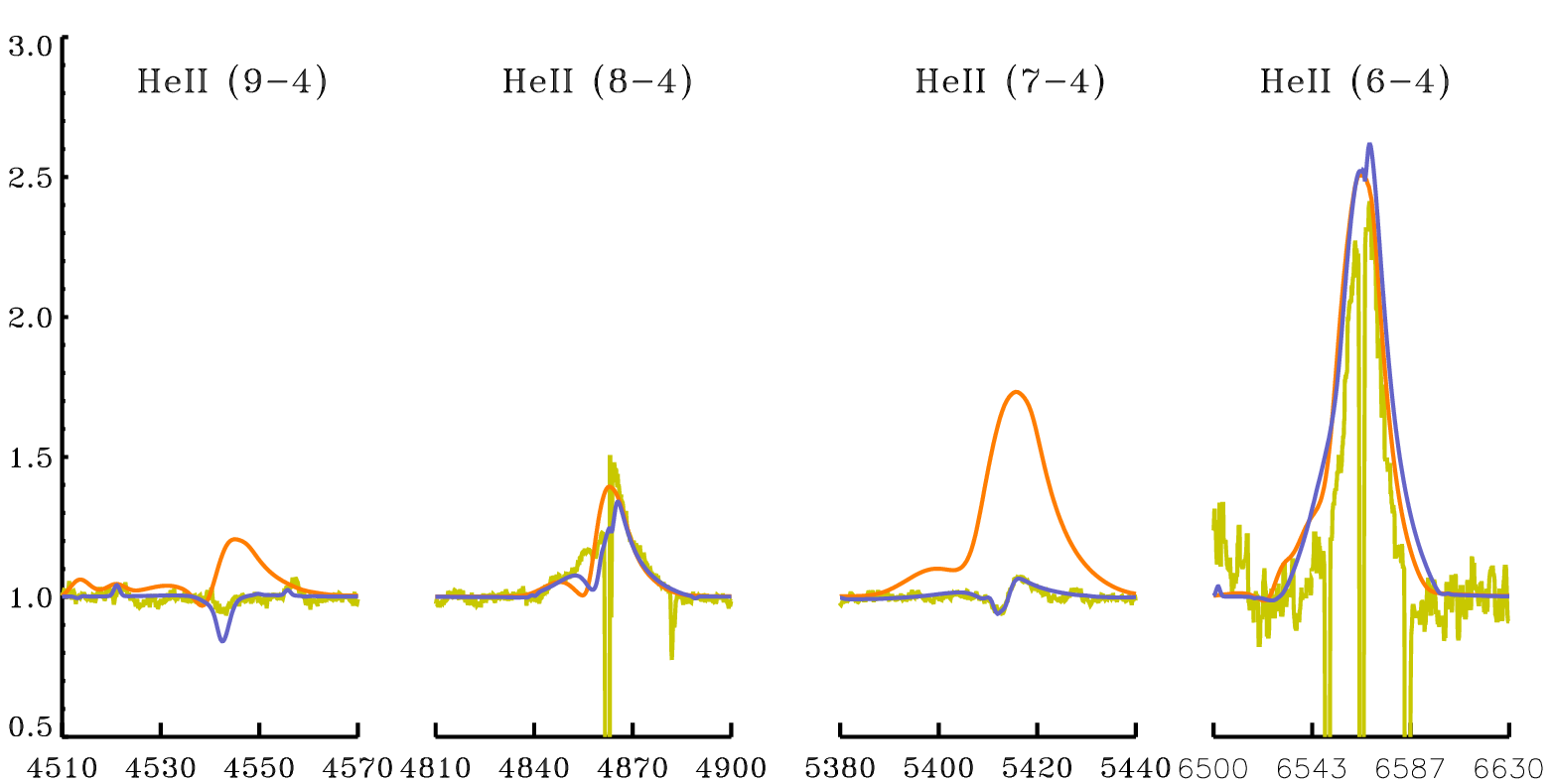}
\caption{\ion{H}{1} and \ion{He}{2} lines from transition $n-4$. The orange
line represents the model with no hydrogen. The blue line represents the model 
with He/H = 0.10. The observations are represented with the green line. Clearly, the H-poor model overestimates the intensities of the lines with $n$ odd.} \label{fig_HHe}
\end{figure}

	The helium content of the wind is one of the fundamental parameters which has to 
be determined at least roughly beforehand. The central star of NGC 6543 has been found to 
be \lq\lq Hydrogen-poor" \citep{1986A&A...169..227B, 1996ASPC...96..141D}, 
so we tested this assumption as the first step 
of our analysis. Due to the similarity of \ion{He}{2} 
and \ion{H}{1} ions, the \ion{He}{2} lines arising from the transition $4-n$ with $n$ even have 
wavelengths close to the 
wavelengths of the \ion{H}{1} lines of the Balmer series. The lines with $n$ odd 
fall between the Balmer lines and are not affected by the hydrogen component. When the wind is 
hydrogen-poor, the 
intensities of the lines with $n$ odd and even decrease smoothly 
with increasing $n$. On the other hand, if the abundance of \ion{H}{1}  is not 
negligible, the intensities of the 
Balmer lines are higher than the \ion{He}{2} lines with $n$ odd 
\citep{1973IAUS...49...15S}. From Fig. \ref{fig_HHe}, it is obvious 
that the central star of NGC 6543 cannot be H-poor. A series of models suggest 
that the He/H ratio should be around solar. We adopt N(H)/N(Ne) = 0.10
in the following analysis.

\subsection{Stellar parameters}

The model of a stellar wind depends upon a large number of parameters. In addition 
to the effective temperature and the
gravity, which describe a plane parallel, static atmosphere, a wind requires 
parameters describing the velocity, the mass loss rate,
 and a parameter describing the clumpiness of the wind. The task of fitting all of the parameters is difficult. We 
proceeded by fixing some of the parameters while fitting the others and then 
iterated, changing the parameters to be fitted.

The parameters describing the star are not independent. If we describe the star 
as an opaque nucleus 
with radius $R_*$ and temperature $T_*$,
then its luminosity is $L = 4 \pi \sigma R_*^2 T_*^4$. It is known 
\citep{1989A&A...210..236S} that 
models with the same temperature $T_*$, velocity law and the same transformed radius
\begin{equation}
R_t = R_* \left(\frac{V_\infty/2500}{\dot M/10^{-8}}\right)^\frac{2}{3}, 
\nonumber
\end{equation}
have very similar emitted spectra. Substituting the radius $R_*$, one obtains a 
scaling rule for the 
mass loss rate if a different luminosity is necessary
\begin{equation}
\frac{\dot M_1}{\dot M_2} = \left(\frac{L_1}{L_2}\right)^{\frac{3}{4}}. 
\label{eq_mdot_l}
\end{equation}

We describe the velocity of the wind using the standard velocity law given by
\begin{equation}
V(r) = V_\infty (1.0-R/r)^\beta,
\end{equation}
fitted to a hydrostatic atmosphere with a gravity $\log{g}$, as described in 
\citet{2003ApJ...588.1039H}.
The terminal velocity $V_\infty$ was set to $1340$ km/s as determined from the 
blue wing of the \ion{P}{5} 1118/28 doublet. The wind of NGC 6543 is not simple. Strong 
lines of different ions with P Cyg profiles require different terminal velocities. 
Ions with higher ionization potential tend to have P Cyg lines with a more 
extended blue wing. In addition, \citet{2007MNRAS.tmp..882P} showed that the
wind variability is present in the absobtion component of \ion{P}{5} 1118/28\AA \  doublet but
\ion{O}{6} 1032/38\AA \ is  more stable. The formation of the strong P Cyg lines is obviously complicated and there is no easy explanation for their behavior. 
We avoided this problem by excluding lines with strong P Cyg profiles from our analysis. We 
account for the presence of different blue wings of the P Cyg lines by increasing the error in $V_\infty$ to $200$ km/s. With the velocity 
law fixed, we continued with the determination of the temperature. 

	There are lines of several elements with two or more consecutive ionization 
stages which, in principle, could be used as temperature indicators. Unfortunately, it is not 
easy to use any of them in practice. The \ion{O}{4} 1339/41 \AA \  doublet and \ion{O}{5} 1371 
\AA \  show different terminal velocities.
In addition, the \ion{O}{5} 1371 \AA \  line is heavily blended with iron lines 
and its intensity strongly depends upon the iron composition. The same is true for the \ion{N}{4} and 
\ion{N}{5} lines. Instead of using ion ratios, we selected several temperature-sensitive features in the spectrum 
and searched for a temperature that reproduces most of them. These features usually depend upon the mass loss 
rate also, so one needs to determine the correct combination of \ $T_{eff}$ and $\dot M$ that reproduce the measured 
fluxes simultaneously. To do that, we ran a grid of models covering temperatures from $60$ to $75$ kK and mass 
loss rates from $2.8\times 10^{-8}$ to $7.8\times 10^{-8}$ M$_\odot$/yr, setting the clumping 
factor to $f = 0.1$ (see below) and the luminosity to $L=5200$ L$_\odot$ 
\citep[as in previous work;][]{1996ASPC...96..141D,2006IAUS..234..401G}. The
temperature limits were set by the \ion{C}{3} and \ion{P}{5} lines. Below $60$ kK
the \ion{C}{3} 1179 \AA \ line is too strong. Above $75$ kK the \ion{P}{5} 
1118/28 \AA \ doublet is too weak.
Once the grid is calculated, one can construct a surface for the value of any 
parameter measured in the 
synthetic spectra as a function of both temperature and $\dot M$. The observed 
value of the parameter 
defines an isoline on that surface. If the isolines defined by several 
parameters are drawn on 
the same coordinates ($T_{eff},\dot M$), their crossing point defines 
the combination of $T_{eff}$ and $\dot M$ which reproduces all of the parameters simultaneously. In 
practice, the isolines do not cross in a point, but rather in a region, whose size defines the error in 
the parameters.  We ran grids for the velocity law setting the parameter $\beta$ equal to 1, 2 and 3. The 
isolines 
cross in a smaller region for $\beta = 2.0$, so we used that value in the subsequent analysis. 

\begin{figure}
\plottwo{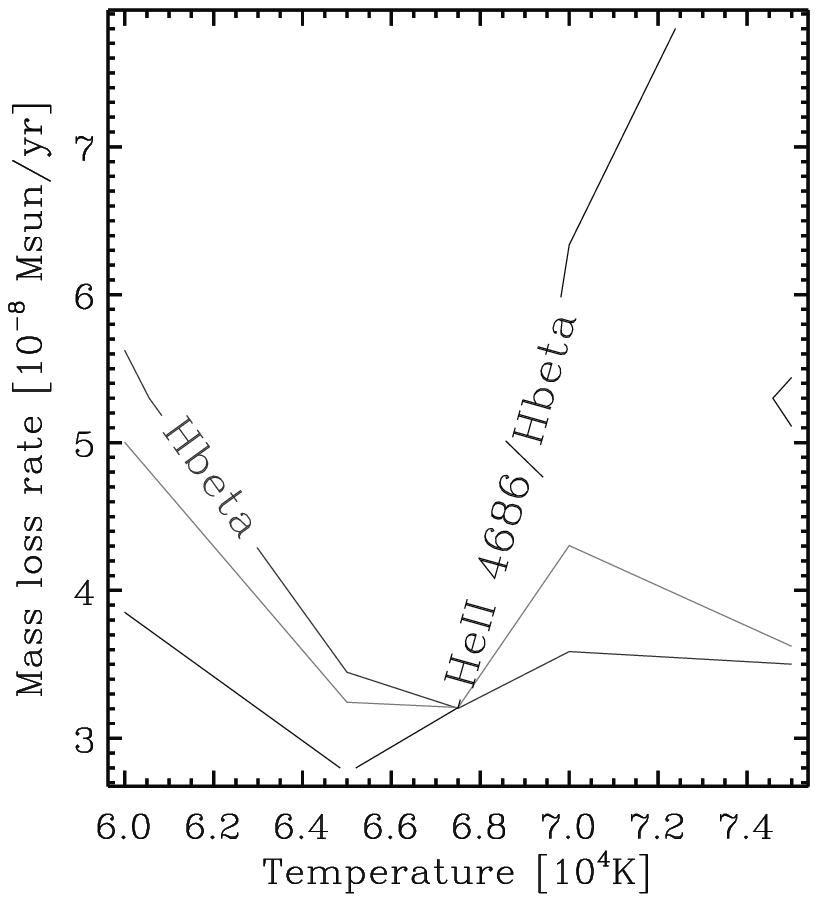}{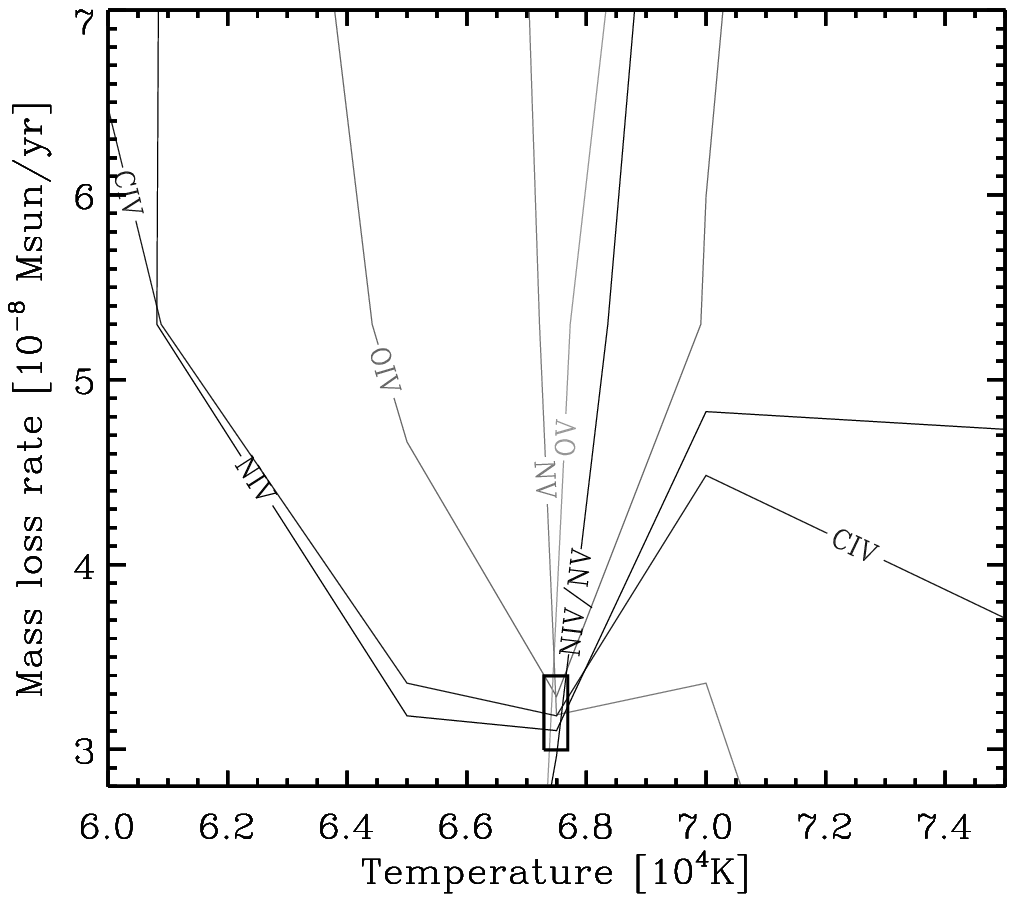}
\caption{Contour lines of the equivalent width of the lines from several ions plotted on a grid of 15 models (see text). Left panel: 
H$_\beta$, \ion{He}{2} 4686 and I(H$_\beta$)/I(\ion{He}{2} 4686) 
(dotted line). Right panel: \ion{C}{4} 5802, \ion{O}{4} 3563, \ion{O}{5} 4930, 
\ion{N}{4} 955 and \ion{N}{5} 4944 lines.} \label{fig_HHegrid}
\end{figure}

\begin{figure}
\plotone{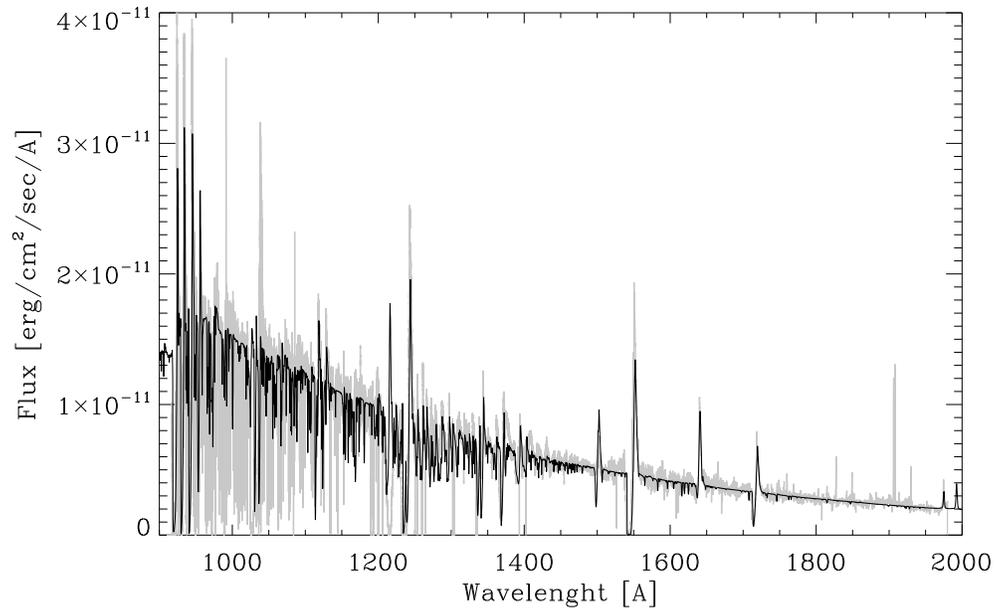}
\caption{Comparison of the observed absolute UV flux (gray) with that predicted by the model (black), adopting a reddening of $E(B-V) = 0.025$\,mag and a distance of 1 kpc}
\label{fig_redening}
\end{figure}

The left panel of Fig.~\ref{fig_HHegrid} shows the isolines for the intensity of \ion{He}{2} 4686\AA \ , $\mathrm H_\beta$,  and their ratio. 
The right panel of the same figure shows several other diagnostics. All of the isolines cross in the vicinity of
$T = 66750\pm 500$ K and $\dot M = 3.2\pm0.05 \times 10^{-8}$ M$_\odot$/yr. We 
adopted these parameters for further modeling.
Once the temperature and the mass loss rate are determined, the absolute flux 
calculated by the model can be compared 
with the observed UV spectrum. The difference provides an estimate of the 
extinction. We used the $R$-dependent Galactic 
extinction curve from \citet{1999PASP..111...63F} for $R = 3.1$. The best fit of 
the line-free regions of the
spectrum is given by $E(B-V) = 0.025\pm0.005$\,mag (Fig.~\ref{fig_redening}) which 
is close enough to the value $E(B-V) = 0.07$\,mag used by 
\citet{2003A&A...406..165B}. The
error in the reddening is the formal error of the fit. The actual error is 
larger, due to its dependence upon the model parameters. 
The scaling factor between the observed UV spectrum and the calculated continuum 
implies a distance of $1.81 \pm 0.05$  kpc, 
where the error reflects the error in the reddening. \citet{1999AJ....118.2430R} 
determined the distance to NGC 6543 of $\sim 1.0$ kpc. To get results in agreement with this 
distance, we scaled down the luminosity to $L = 1585\,L_\odot$
and the mass loss rate  $\dot M = 1.86 \pm 0.3 \times 10^{-8}$ M$_\odot/yr$, according to (\ref{eq_mdot_l}), where the error 
also reflects the uncertainty in
$V_\infty$. We stress that the model itself does not provide 
a distance due to the degeneracy between the luminosity and the mass loss rate mentioned above. 

Finally, we have to address two additional parameters. First is the stellar mass. 
The assumed luminosity $L = 1585\,L_\odot$ and the adopted temperature implies 
a value of $\log{g} = 5.3$ if the stellar mass is $0.6$ M$_\odot$. The mass of the star can 
vary by a factor of 1.3,
 which leads to changes in $\log{g}$ of $\pm 0.12$. We did test runs
with higher and lower gravity. There were no significant changes in the emitted 
spectrum, so we think that a different 
value of $\log{g}$ (necessary for a different luminosity) would not change the 
conclusions made in the next section. On the other hand,
the wings of the hydrogen lines are well reproduced (Fig.~\ref{fig_4100}), which
indicates that the adopted value of $g$ is adequate.

Another parameter that can change the emitted spectrum is the wind clumpiness. 
The winds of massive stars are clumpy,
as deduced from the electron scattering wings of strong lines \citep{2005ASPC..332..215H}. The 
same lines in the spectrum of NGC 6543 are
not strong enough to permit us to determine the clumping factor $f$. In general, 
models with the same
$\dot M/\sqrt{f}$ produce similar spectra. We ran test models with $f = 0.05, 0.1, 0.5$ 
and $1.0$ with corresponding
changes in the mass loss rates. Only a few lines changed, between them the
strong \ion{C}{4} 5801/12 doublet. Since we did not use these lines in the composition analysis, we set the clumping factor to $f = 0.1$
which is similar to the value $f=0.08$ obtained by \citet{2007MNRAS.tmp..882P}.

\subsection{Chemical composition}

Once the parameters of the star are determined, the composition of the elements
is determined by comparing the predicted
and observed line fluxes.  As mentioned above, the strong resonance lines are
poorly reproduced and are not useful for
abundance determinations. The following analysis is based upon weaker lines that 
are formed deep in the wind and in a smaller volume.
These lines are better reproduced by the model and we expect that they are less
sensitive to the perturbations of the wind structure
and geometry. The model reproduces most of the lines correctly,
but we could not reproduce everything. Some of the observed features are missing
in the model spectrum. One has to keep in mind that the model presented here
includes most of the important physics and chemistry, but not all of them. Given this consideration, a comparison between observations and 
models with different abundances yields the
errors presented in Table~\ref{tab_comp}.

\begin{figure}
\plotone{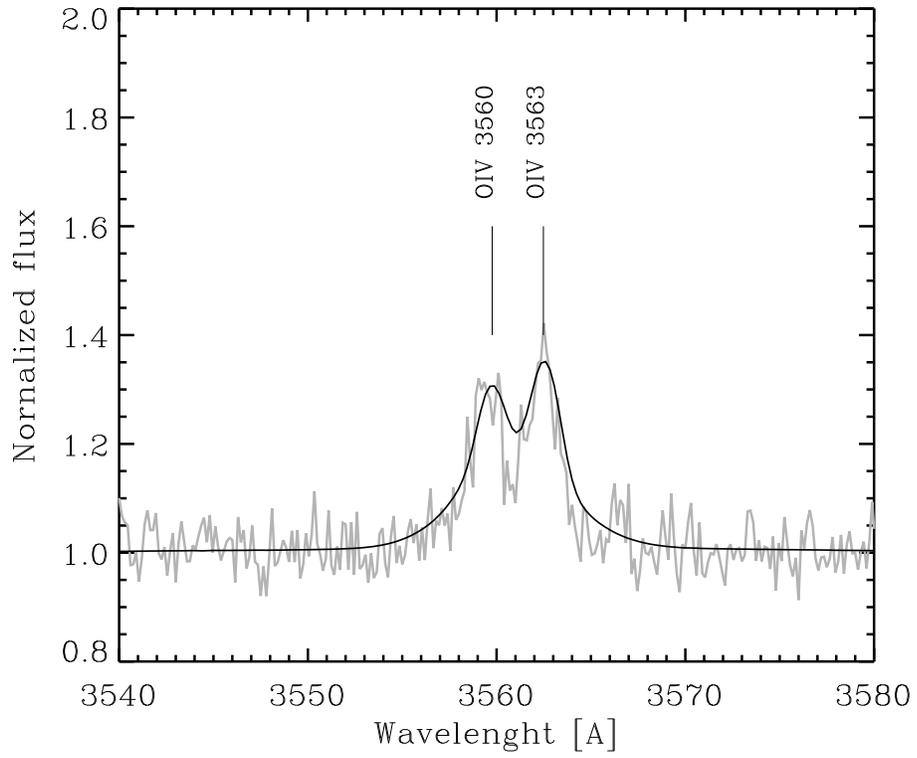}
\caption{The fit to the \ion{O}{4}  3560/63 \AA \  doublet.} \label{fig_OVI}
\end{figure}

\begin{figure}
\epsscale{1.0}
\plotone{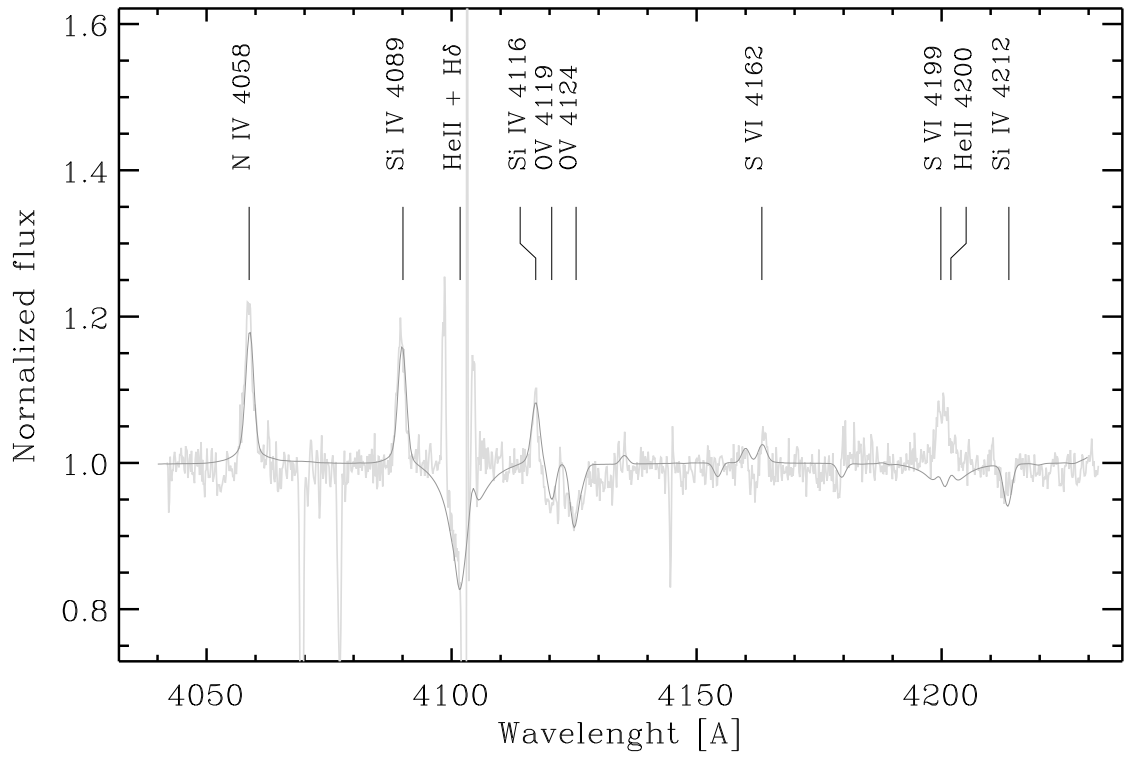}
\caption{Comparison of the observed spectrum and the model. The \ion{Si}{4} 
lines are reproduced with an enhanced silicon abundance.} 
\label{fig_4100}
\end{figure}

\begin{figure}
\plotone{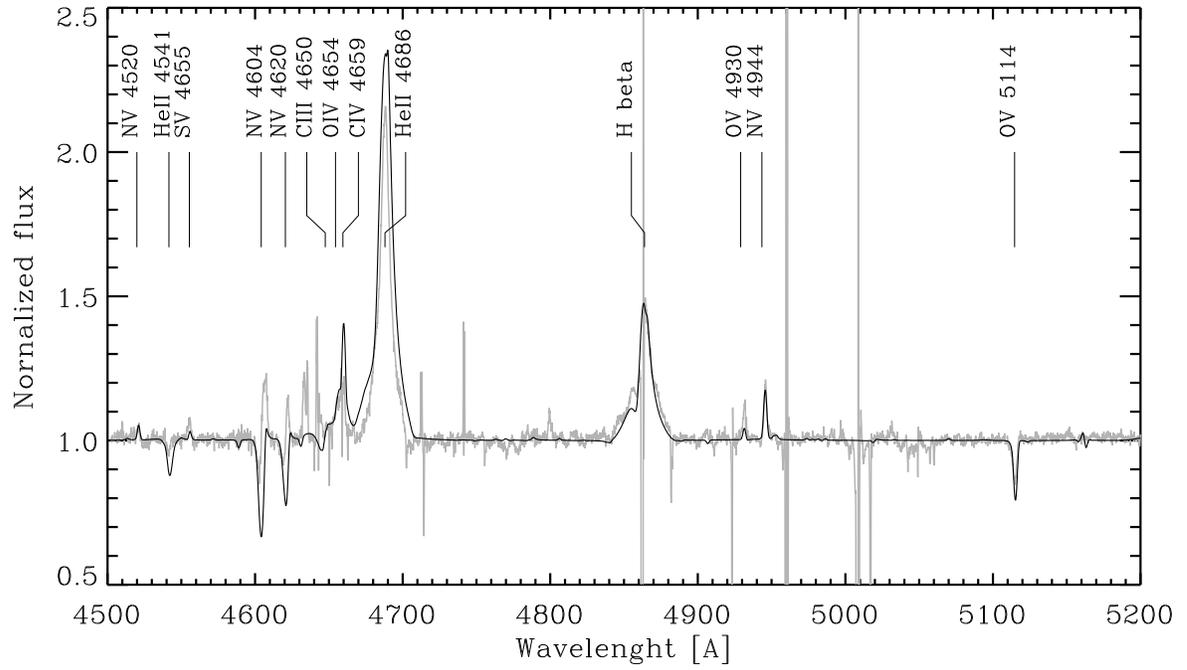}
\caption{Comparison of the observed spectrum and the model. Most of the 
lines are well reproduced, except \ion{N}{5} 4604/20, though note that \ion{N}{5} 4944 is reproduced.} \label{fig_4686}
\end{figure}

\begin{figure}
\plotone{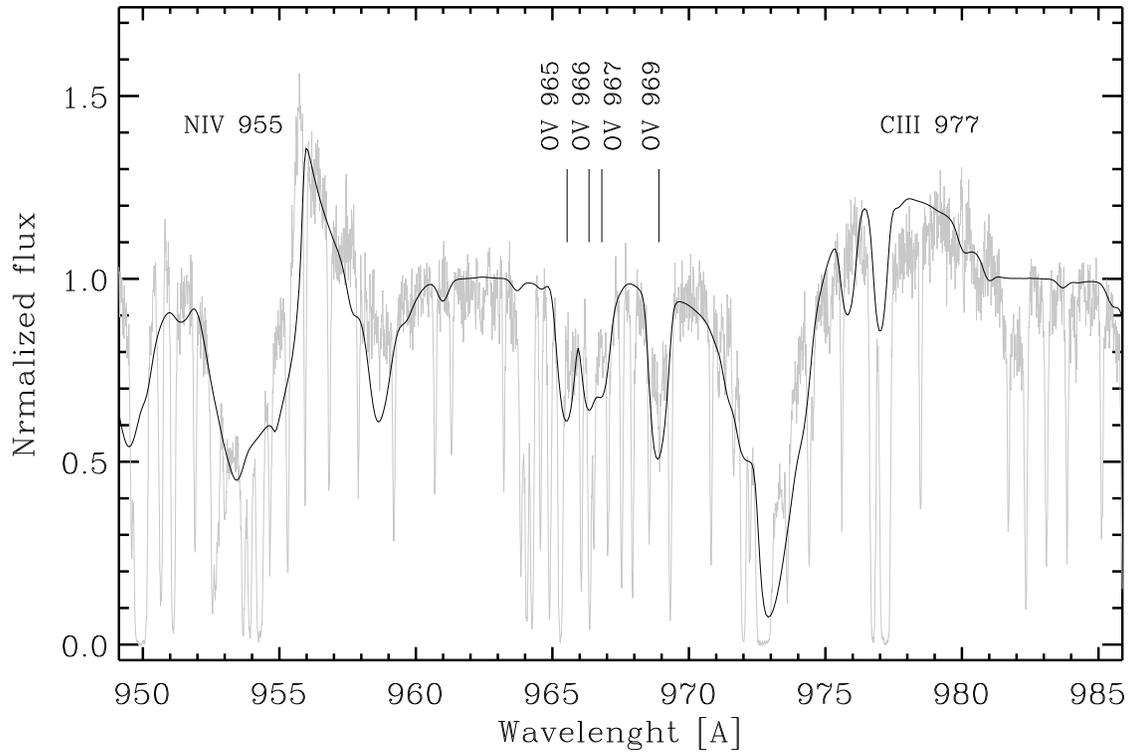}
\caption{Comparison of the observed spectrum and the model in the FUV.} 
\label{fig_965}
\end{figure}

Oxygen is found mainly as \ion{O}{4} and \ion{O}{5}. We determined its
abundance using the \ion{O}{4} doublet at 3560/63 \AA.
This line is blend-free and has good S/N
(Fig.~\ref{fig_OVI}). The line intensity is reproduced with an oxygen abundance
of $12+\log N($O$)/N($H$)=9.02$\,dex. As an additional criteria to check the O abundance,
we used the \ion{O}{5} 4119 \ \AA, \ion{O}{5} 4123 \ \AA, and  \ion{O}{5} 5114 \AA \ lines (Figs. \ref{fig_4100} and \ref{fig_4686})
which were well reproduced.
In addition, there are a few \ion{O}{5} lines around 965\ \AA \
(Fig.~\ref{fig_965}) that are overestimated which points to an 
overabundance of \ion{O}{5} in the model with respect to the observations. This is
in agreement with the additional ionization of \ion{O}{5} (see below) and we 
adopt the abundance derived from \ion{O}{4} lines.

In addition to the \ion{O}{4} and \ion{O}{5} lines, the FUSE spectrum shows a
very strong \ion{O}{6} 1032/38\AA \  resonance doublet.  The model
with the temperature and mass loss rate adopted above underestimates these lines
by an order of magnitude. One needs a temperature above 90 kK and
a higher mass loss rate to reproduce this line. It is believed that the
\ion{O}{6} 1032/38\AA \  doublet is formed by the ionization
of \ion{O}{5} by X-ray radiation generated by internal shocks in the wind \citep{1981ApJ...250..677C}. In support of this hypothesis, only
\ion{O}{6} 1032/38\AA \  is observed and other strong \ion{O}{6} lines, such as the
3811/34 \AA \ doublet, are not present in our spectra.  Test models with
higher a temperature, which reproduce \ion{O}{6} 1032/38 \AA, also show an observable \ion{O}{6} 3811/34 \AA \  doublet.  These arguments
imply an additional ionization source in the external part of the wind, which does not affect the internal, denser,
part of the wind. The density in the highly-ionized part of the wind is very low meaning that only the resonance lines have enough opacity to
form an observable feature, i.e., producing \ion{O}{6} 1032/38 \AA\ , but not \ion{O}{6} 3811/34 \AA.  One could speculate that the
additional ionization source is the diffuse X-ray plasma interior to the nebular shell. Fitting the  \ion{O}{6} 1032/38\AA \  lines with
shock-generated or other X-ray emission is beyond the scope of this paper.  The
abundance analysis based upon weak lines of low ionization stages
should not be affected by the X-ray ionization and should be correct.

\begin{figure}
\plottwo{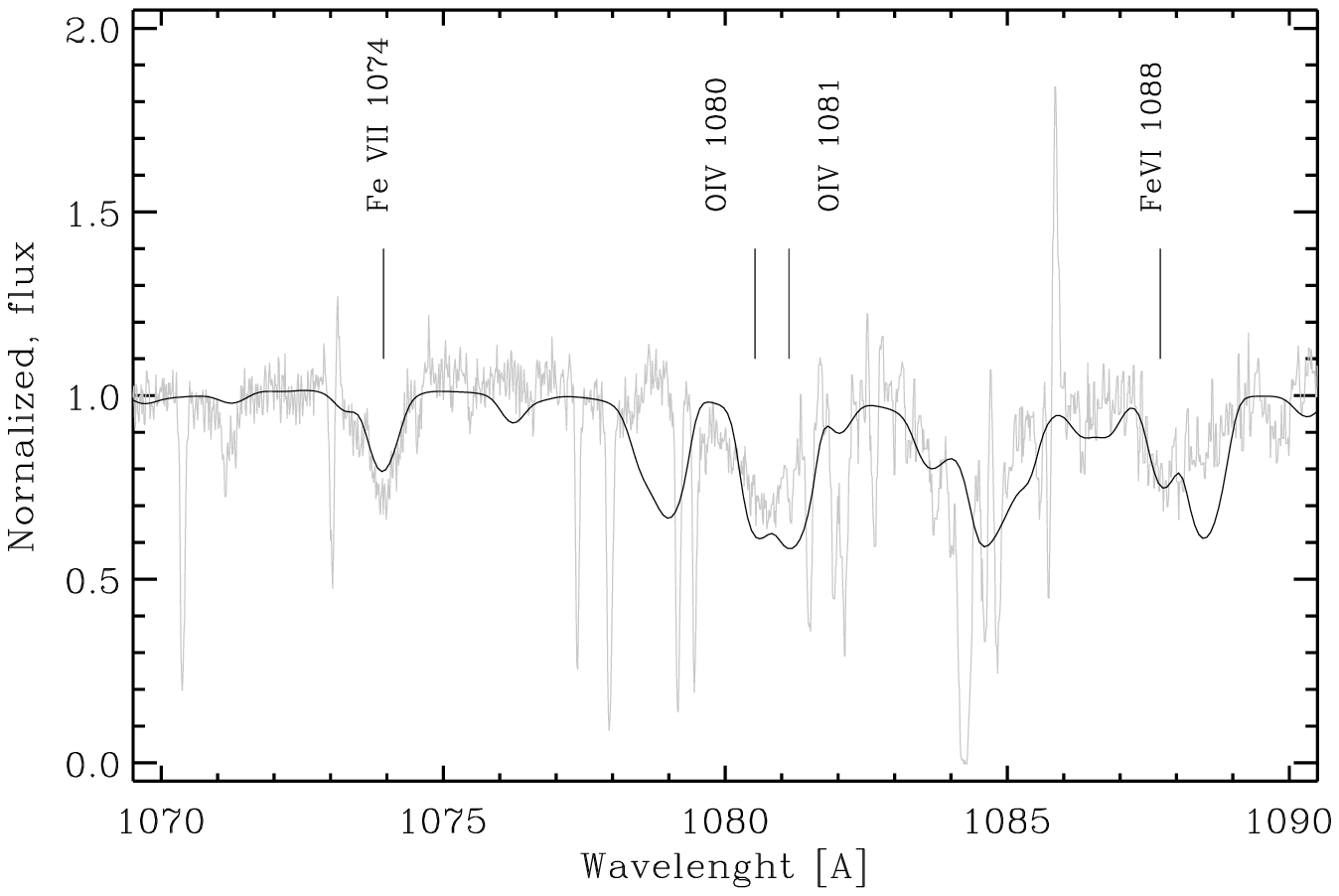}{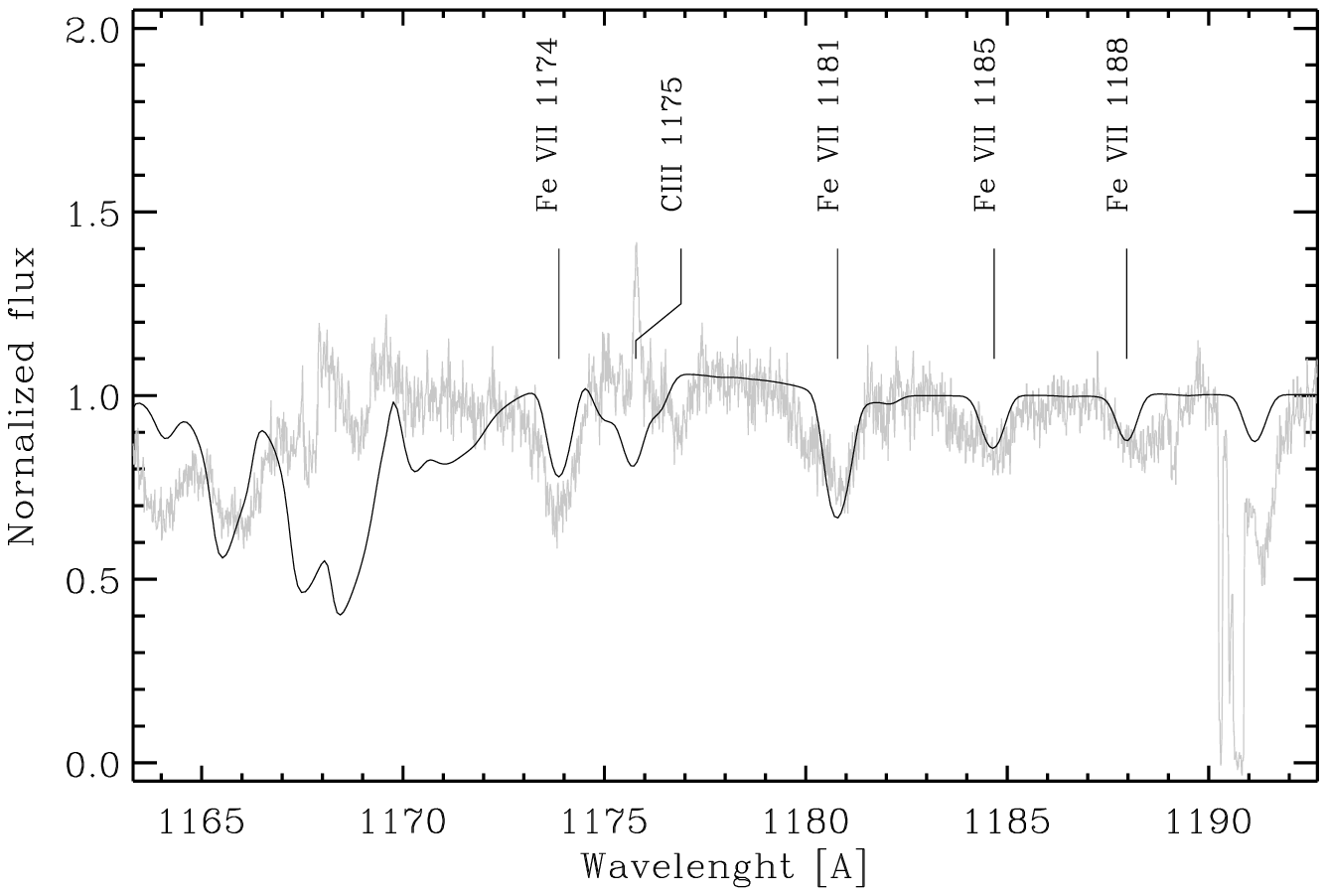}
\caption{Lines used for measuring the iron composition. A variety of lines are marked.} \label{fig_FeVII}
\end{figure}

Carbon is represented by two ionization stages, with \ion{C}{4} being the
dominant one. Lines of \ion{C}{3} and
\ion{C}{4} are observed in the FUSE, IUE, and optical regions
of the spectrum. The \ion{C}{4} resonance doublet 1548/50 \AA \ behaves
similarly to \ion{O}{6} 1032/38\AA \ . It has a higher $V_\infty$ and a larger
observed intensity than the model predicts. We determined the carbon
composition mainly from the
\ion{C}{3} 1176 \AA \  and 977 \AA \  lines (Figs. \ref{fig_FeVII} and
\ref{fig_965}).
These two lines, together with \ion{C}{4} 1169 \AA,  were used as a temperature
diagnostic. Our final estimate is $12+\log N(\mathrm C)/N(\mathrm H)=9.03$\,dex.

Nitrogen is mainly in the form of \ion{N}{5} and, like \ion{O}{6} and \ion{C}{4}, it is also apparently strongly affected
by the X-ray ionization. We estimate its
composition using the \ion{N}{4} 955\AA, \ion{N}{4} 1718\AA,
\ion{N}{4} 4058 \AA,
and \ion{N}{5} 4944 \AA \  lines (Figs. \ref{fig_965}, \ref{fig_4100}, and \ref{fig_4686}, respectively).  The lines are
reproduced with $12+\log N($N$)/N($H$)=8.36$\,dex.

The \ion{Si}{4} resonance doublet 1398/1402 is usually present in many objects.
Due to the high temperature of the star, the lines are weak and heavily blended with \ion{Fe}{6} lines. Instead of using the
resonance doublet, we estimated the silicon abundance
using the \ion{Si}{4} 4089 \AA \ and 4116 \AA \  lines (Fig.~\ref{fig_4100}).
The lines are reproduced with $12+\log N($Si$)/N($H$)=8.19$\,dex, a factor
of five higher than the solar value. The model with this abundance reproduces
several other \ion{Si}{4} lines as well.
We have to stress that the \ion{Si}{4} ion is not the dominant stage of silicon at a temperature of $\sim$ 70 kK. Only 0.2\% of
the silicon is in \ion{Si}{4}. Therefore, the estimated overabundance depends
strongly on the atomic data and it is highly uncertain. On the other hand, \citet{2007A&A...467.1253J}
found similar silicon overabundance in the central star of M1-37. Silicon overabundance is an unexpected 
result which has no explanation for the moment. 

In addition to the above mentioned elements, the model contains sulfur and
phosphorus. The sulfur line \ion{S}{5} 1503 \AA \ was fitted with
$12+\log N(\mathrm S)/N(\mathrm H)=7.57$\,dex. We checked that this abundance also reproduces the 
\ion{S}{6} 4162 (Fig~\ref{fig_4100}) line well.
The phosphorus lines \ion{P}{5} 1118/1128 are well-fitted with an abundance of $12+\log N(\mathrm P)/N(\mathrm H)=5.53$\,dex.

\begin{figure}
\plotone{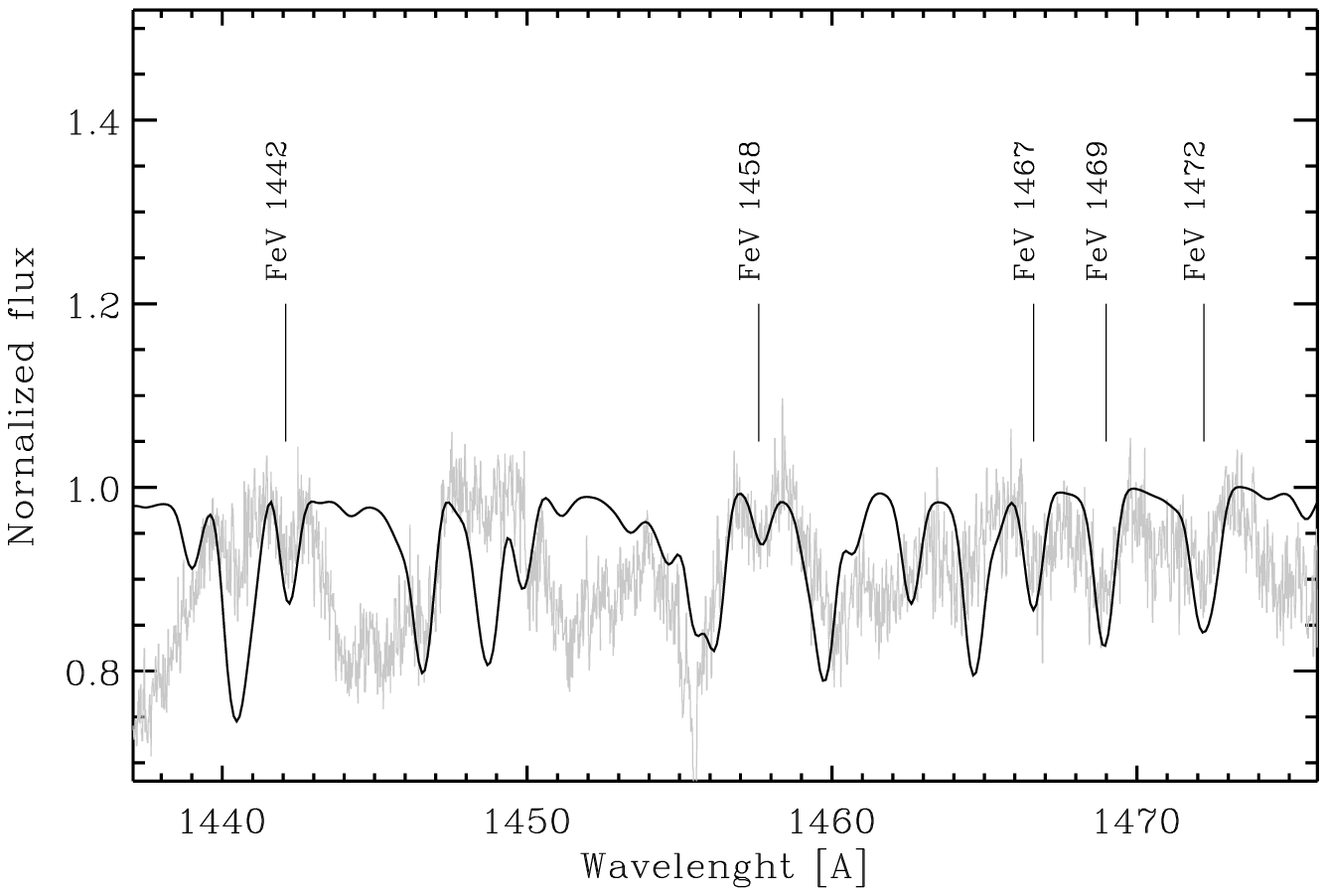}
\caption{Lines used for measuring the iron composition. The \ion{Fe}{5} lines are marked.} 
\label{fig_FeV}
\end{figure}

Finally, we analyzed the iron content. A first look at the spectrum reveals 
strong iron lines between 1250 \AA \  and 1450 \AA. The
 presence of these lines rules out a significant depletion of iron 
\citep{2006IAUS..234..401G}, but the precise modeling showed
 that these features are not very sensitive to variations in the iron abundance. A similar result was reported by
 \citet{2007ApJ...654.1068M} for \BD. Therefore, we decided to apply the 
classical curve of growth method. Once we had 
the wind parameters and composition fixed, we ran several models with different 
iron abundances, starting from the solar value and
 finishing with 1/10 of the solar value. We selected about 10 absorption lines 
mainly \ion{Fe}{5} and \ion{Fe}{7} (Figs.~\ref{fig_FeV} 
and \ref{fig_FeVII}) that showed the
 strongest variation with composition. The equivalent widths of the lines both 
in the observed spectrum and in the models were
 measured using a Gaussian fit. Using the curve of growth method, we derived the 
Fe abundance for each line. The median value yields 
$12+\log N($Fe$)/N($H$)= 7.24$\,dex which is $\sim 0.2$\,dex lower than the solar value.  

The model also includes lines of argon, neon and nickel. The abundances 
of these elements were set to their solar values and were not determined by line fitting.

Our elemental abundances for the central star's wind are presented in Table~\ref{tab_comp} along with those for the Sun and the Orion nebula. 
Note that the Ne and Ar abundance determinations for the Orion nebula
are more accurate than those for the Sun. The uncertainties in the abundances for NGC 6543, except in the case of iron, are derived from 
models with increased and decreased abundances. The uncertainty reflects the changes in the abundances that 
produce changes greater than the noise in the data. The uncertainty in the iron abundance is
the standard deviation of the abundance derived for each of the lines analyzed. 

Finally, we stress that our model of the central star's wind is undoubtedly only one of several possible solutions. Several combinations of parameters produce similar spectra and therefore increases the uncertainty in the stellar parameters, which is reflected by the errors in the derived abundances.  In addition, the observed spectrum shows features that are not present in our model.  Nonetheless, the precision of the chemical composition presented here is sufficiently robust for the purposes of this paper. A more refined model, dealing with the formation of \ion{O}{6} lines, the identification of the observed features and the difference in the wind velocity shown by different resonance lines will be treated in a forthcoming paper.  A final conclusion on the evolutionary status of the object and the interrelation between its components (star, nebula, X-ray emitting gas) require a self consistent model including the central star and the nebula (Morisset \& Georgiev, in preparation). 

\section{Relationship between the stellar and nebular abundances}

In most planetary nebulae, the abundances for heavy elements derived from recombination and collisionally-excited lines do not agree.  Explanations for this result depend fundamentally upon large temperature variations that cannot be explained by simple photoionization models \citep[e.g.,][and references
therein]{2006IAUS..234..219L,2006IAUS..234..227P}. The temperature variations 
explain the large differences between the chemical abundances derived from recombination lines and and those derived from collisionally excited lines when a constant temperature is used.  The difference between both types of abundances is called the abundance discrepancy factor, $ADF$. There are two different ideas
to explain the $ADF's$: (a) the presence of temperature variations in a
chemically inhomogeneous medium \citep[e. g.][and references 
therein]{1990A&A...233..540T,2006IAUS..234..219L}, and (b) the presence of temperature variations due to other causes in a chemically homogeneous medium
\citep[e. g.][ and references therein]{2006IAUS..234..227P}. 

According to the two-abundance nebular model by Liu and collaborators, PNe present two components: (a) a low density component that has most of the mass and is relatively hot, emits practically all the intensity of the H lines and of the forbidden lines in the visual and the UV, as well as part of the intensity of the He~I lines, and (b) a high density component that has only a small fraction of the total mass, is relatively cool, H-poor, rich in heavy elements, emits part of the He~I and all of the recombination lines of the heavy elements, but emits practically no H nor any collisionally-excited lines from heavy elements.  Chemically inhomogeneous nebulae can be produced by H-poor stars that eject H-poor material into H-rich nebulae.  That is the case in A30 and A78 \citep{1979PASP...91..754J,1980Natur.285..463H,1983ApJ...266..298J,
1988A&A...191..128M,2003MNRAS.340..253W}. This type of situation
might occur in those cases where the central star has an H-poor atmosphere.
According to \citet{2000A&A...362.1008G} about 10\% of the central stars
of PNe are H-poor. Studies based upon the Sloan project find similar results: based upon 2065 DA and DB white dwarfs, \citet{2004ApJ...607..426K} find that 1888 are non magnetic DAs and 177 are non magnetic DBs. Therefore, we conclude that about 10\% of Galactic PNe have a H-poor central star, and might show He, C, and O rich inclusions in their expanding shells.  We consider it unlikely that PNe with H-rich central stars would contain significant amounts of H-poor material in their associated nebulae.

\subsection{Nebular collisional and recombination abundances}

In Table~\ref{tta:Col}, we present abundances derived from collisionally-excited lines from different studies. In the last column, we present our adopted values based upon the values presented in the previous columns.  For O$^{++}$, the value taken from \citet{2004MNRAS.351.1026W} is based only upon their [\ion{O}{3}] $\lambda 4959$ determination, since the $\lambda\lambda$ 52 and 88 $\mu$m lines have a different dependence upon the temperature than the [\ion{O}{3}] $\lambda\lambda 4959$ and $5007$ lines.  Also, the presence of density variations affects the O abundance determination based on the $\lambda\lambda$ 52 and 88 $\mu$m lines.

In Table~\ref{tta:Rec}, we present abundances derived from recombination lines
from different studies.  In the last column we present our adopted values. The $N$(He)/$N$(H) ratio was derived by us from helium recombination lines and using more recent atomic data than those used in other studies.  The detailed abundance determination is based upon maximum likelihood method, MLM, and is 
discussed in section~\ref{sec_t2}. For C, the adopted value is just the average 
of the three determinations.  For O$^{++}$, the adopted abundance value is based
upon the determination of \citet{2004MNRAS.351.1026W}, but with two modifications: (a) we weighted the contribution of each multiplet according to its effective recombination coefficient, and (b) we eliminated multiplet V12 that might be contaminated by other emission lines.  Our determination yields
$N$(O$^{++}$)/$N$(H$^+$) = 1.42 x 10$^{-3}$.  Note that V1, the brightest
multiplet, yields 1.18 x 10$^{-3}$ for this ratio.  Adopting an ionization correction factor of 1.09 derived from the [O~{\sc{ii}}] $\lambda\lambda$ 3726,3729 lines and the $N$(O$^{++}$)/$N$(H$^+$) recombination ratio, we obtain an oxygen abundance $12 + \log N(\mathrm O)/N(\mathrm H) = 9.19$\,dex. At first sight the Ne/H ratio derived from recombination lines seems to be high (see Table 4), but the values for $\log$Ne/O values for the CELs and RLs amount to -0.52 and -0.54 dex, respectively, in good agreement with the values derived for a large number of PNe by \citet{1977RMxAA.2....181T} and \citet{1994MNRAS.271..257K}, who find an average $\log \mathrm{Ne}/\mathrm O$ ratio of -0.59\,dex.

\subsection{$t^2$ value determinations \label{sec_t2}}

In Table~\ref{tta:Tem}, we present various temperature determinations for NGC 6543. $T$(Bac) is the temperature determined from the intensity ratio of the Balmer continuum to a Balmer line.  $T$(He~{\sc{ii}}) is the temperature derived from the He I recombination lines (see next subsection).  $T$[O~{\sc{iii}}] and $T$[O~{\sc{ii}}] are the temperatures derived from the ratio of the auroral and nebular line intensities for the corresponding ions. $T$([O~{\sc{iii}}],[O~{\sc{ii}}]) is the representative temperature for the forbidden lines, where we are assuming that 92\% of the emission originates in the O$^{++}$ zone and 8\% in the O$^+$ zone. The differences among the various temperatures imply the presence of temperature variations within the observed volume.  Also, the differences between the collisional and recombination abundances, the $ADF$ values, for C, N, O, and Ne also imply the presence of temperature variations.

To reconcile the differences among the various temperatures and between the
collisional and recombination abundances it is possible to characterize the 
temperature structure by an average temperature, $T_0$ and a mean square 
temperature variation, $t^2$. These quantities are given by
\begin{equation}
T_0\left(N_e,N_i\right) = \frac{\int 
T_e\left(r\right)N_e\left(r\right)N_i\left(r\right) dV}{\int 
N_e\left(r\right)N_i\left(r\right) dV}
\end{equation}
and
\begin{equation}
t^2 = \frac{\int \left(T_e-T_0\right)^2 N_eN_i dV}{T_0^2 \int N_eN_i dV}
\end{equation}
where $N_e$ and $N_i$ are the electron and the ion densities, respectively, of the observed 
emission line and $V$ is the observed volume
\citep{1967ApJ...150..825P}.

Under the assumption of chemical homogeneity to derive a $t^2$ value, we need two independent temperature determinations,  and the temperature dependence of the line or continuum intensities used to derive the temperature.  It is also possible to derive a $t^2$ value for a particular ion by reconciling the abundances derived for this ion derived from the intensities of collisional and recombination lines \citep[][and references therein]{2004ApJS..150..431P}.  The  
result is correct even in the presence of chemical inhomogeneities.

In Table~\ref{tta:tvalues}, we present five independent $t^2$ determinations. The first two were obtained from the comparison of  two temperatures representative of the whole observed volume, and the other three were derived under the assumption that the collisional and the recombination abundances had to be the same.  The temperature dependencies of the recombination lines of C$^{++}$, O$^{++}$, and Ne$^{++}$, needed to determine the $t^2$ values were obtained from \citet{2000A&AS..142...85D}, \citet{1994A&A...282..999S}, and \citet{1998A&AS..133..257K}, respectively.

\subsection{Physical conditions derived from the helium recombination lines}

To obtain $N(\mathrm{He}^+)/N(\mathrm H^+)$ values, we need a set of effective recombination coefficients for the helium and hydrogen lines, an estimate of the optical depth effects for the \ion{He}{1} lines, and the contribution to the \ion{He}{1} line intensities  due to collisional excitation.  We used the hydrogen recombination coefficients from \citet{1995MNRAS.272...41S}, the helium recombination coefficients from \citet{2005ApJ...622L..73P}, with the interpolation formulae provided by \citet{2007ApJ...657..327P}, and the collisional contribution to the \ion{He}{1} lines by \citet{1993ADNDT..55...81S} and \citet{1995ApJ...442..714K}. The optical depth effects in the triplet lines were estimated from the computations by \citet{2002ApJ...569..288B}.

To derive the physical conditions associated with the helium ionized region, we have used a maximum likelihood method, MLM \citep{2000ApJ...541..688P,2002ApJ...565..668P}.   To determine, $N_e$(He~{\sc{ii}}), $T_e$(He~{\sc{ii}}), 
$N({\rm He}^+)/N({\rm H}^+)$, and the optical depth in the \ion{He}{1} 3889 \AA\ line, ($\tau_{3889}$), 
self-consistently, we used as inputs a characteristic density from the forbidden 
line ratios of $N_e = 4000\pm 2000$ cm$^{-3}$ and 13 $I$(He~{\sc{i}})/ 
$I$(H~{\sc{i}}) line ratios observed by \citet[][the
13 He~{\sc{i}} lines are $\lambda\lambda$ 3820, 3889, 3965, 4026, 4387, 4438, 
4471, 4713, 4922, 5876, 6678, 7065, and 7281]{2004MNRAS.351.1026W}.  Each of the 14 observational constraints depends upon $T_e$(He~{\sc{ii}}), $N_e$(He~{\sc{ii}}), $N({\rm He}^+)/N({\rm H}^+)$, and $\tau_{3889}$, each dependence being unique.  Therefore,
we have a system of 14 equations and 4 unknowns.  We obtain the best value for the 4 unknowns and $t^2$ by minimizing $\chi^2$.  In this way, we obtained that $t^2 = 0.035 \pm 0.014$, $T_e$(He~{\sc{ii}}) = 6674 $\pm 559$ K, $N_e$(He~{\sc{ii}}) = 3383 $\pm 2100$ cm$^{-3}$, $\tau_{3889} = 9.4 \pm0.9$, and $N({\rm He}^+)/N({\rm H}^+) = 0.1130 \pm 0.0023$.

\subsection{Comparison of stellar and nebular abundances}

In Table~\ref{tta:SteNeb}, we present the stellar abundances for NGC~6543 based upon the non-LTE model and compare them with the nebular abundances derived from recombination lines (RL) and collisionally excited lines (CEL).  The agreement between the nebular recombination line abundances and the stellar abundances for He, C, and O is excellent, which indicates that the central star is not ejecting H-poor material into the nebula, material that is needed to support the two-abundances nebular model.  From the five independent $t^2$ values, we conclude that the temperature structure for H, He, C, O, and Ne is similar and therefore that the proper heavy element abundances to compare with those of H and He are those derived from recombination lines.  This conclusion is reasonable because the temperature dependence of the recombination lines of the five elements is weak and similar.  Consequently, the presence of temperature variations does not affect the derived abundances. The opposite is the case when comparing collisionally excited lines with recombination lines because their temperature dependence is very different \citep[e. g.][]{1967ApJ...150..825P}. In other words, the $ADF's$ in NGC 6543 are solely due to temperature variations in a medium where H, He, C, O, and Ne are homogeneously distributed.  Our results are in disagreement with the conclusions of \citet{2004MNRAS.351.1026W} who propose 
that this object contains high-density, H-poor inclusions that are rich in helium and heavy elements.

\section{Comparison of the abundances in the stellar wind and the X-ray plasma}

Iron depletion in a hydrogen-rich CSPN is an unexpected result. It is generally believed that H-poor star are also iron-poor, but that H-rich stars have normal iron content \citep{miksaetal2002, heraldbianchi2004a, heraldbianchi2004b, heraldbianchi2004c, stasinskaetal2004}.  As Figs.~\ref{fig_FeV} and ~\ref{fig_FeVII} show, some of the iron lines are well-reproduced, and the equivalent width of others is over-estimated.  As a result, it seems clear that we may reject a solar iron composition for the stellar wind of the central star in NGC 6543. Nonetheless, the depletion of iron is moderate. 

In our previous paper \citep{2006ApJ...639..185G}, we showed that the X-ray-emitting plasma in NGC 6543 is heavily depleted in iron.  This chemical peculiarity allows one to determine the origin of this plasma by comparison with the iron abundances in the stellar wind and the nebular gas.   In other words, are the X-rays emitted by the shocked stellar wind, whose temperature is reduced by some mechanism \citep[for proposed scenarios, see][]{2003ApJ...583..368S}, or are the X-rays emitted by heated nebular gas?  We find that the wind from the central star is depleted in iron, but by a factor of only 1.6 with respect to the solar iron abundance.  The depletion needed to explain the missing [\ion{Fe}{14}] 5303 \AA \ line in the X-ray plasma is at least a factor of 10.  As a result, the X-ray emitting gas cannot be related to the stellar wind.  On the other hand, the nebular gas is iron-depleted. The iron abundance in the nebular gas is estimated to be depleted by a factor of 11 compared to the solar abundance \citep{1999A&A...347..967P}.  Therefore, the hot gas appears to be of nebular origin.  

At least for NGC 6543, therefore, one can interpret the formation of the hot, X-ray-emitting plasma as the result of heating nebular material.  One possible mechanism to accomplish this is thermal conduction, as suggested by \citet{1994AJ....107..276S} and \citet{1996A&A...309..648Z,1998A&A...334..239Z}.  This mechanism also explains the low temperature of the X-ray emitting gas.  In this case, the observed temperature is not directly related to the deposition of the mechanical energy by the wind, but rather to the efficiency of thermal conduction. The cold nebular gas is heated by thermal conduction to the observed temperature of the hot plasma and so emits in diffuse X-rays.  

The same scenario can be extended to at least one other object with diffuse X-ray emission. \citet{2007ApJ...654.1068M} showed that the central star of BD+30$^{\circ}$3639 has an iron abundance that is only moderately depleted. They concluded that the iron forest in the UV is reproduced with an iron content equal to 1/4 of the solar value, but they cannot rule out the solar composition either.  Previously, we have found the X-ray-emitting plasma to be depleted in iron by a factor of 8 \citep{2006ApJ...639..185G}.  This result again implies that the hot gas arises from nebular material rather than from the stellar wind. 

\section{The evolution of the progenitor of NGC 6543}

In addition to the iron abundance in the stellar wind, we derive the abundances of CNO and some other elements that constrain the evolution of the star before it became a planetary nebula. The most important abundance anomaly is the carbon abundance.  Compared to the solar composition, carbon is enriched significantly, while oxygen and nitrogen are almost normal.  In addition, iron is depleted.  These abundances imply that the progenitor star had a mass below 4.0 $M_\odot$, but higher than 1.8 $M_\odot$. The upper limit follows from the normal nitrogen composition. Stars with masses above 4.0 $M_\odot$ experience hot bottom burning and show large nitrogen and helium enhancements \citep{2005ARA&A..43..435H}. On the other hand, carbon is overabundant, which requires a carbon-rich intershell, formed only in stars with masses above 1.8 $M_\odot$. The depletion of iron points to the presence of effective s-processes.  We cannot observe any of the elements heavier than iron and cannot determine their abundances, but the iron depletion is an indirect indicator of these processes. 

\section{Conclusions}

We have obtained a detailed atmospheric NLTE model of the central star of NGC 
6543 that includes a stellar wind and that matches a large number of the observed emission and absorption lines. The main physical parameters of the model are: $T_{eff}= 66750$\,K, $R_{\mathrm{star}}=1.97\times 10^{10}$\,cm, $L = 1585\, L_\odot$, $\log g= 5.3\,\mathrm{cm\, s}^{-2}$, $\dot M = 1.86\times 10^{-8}\,M_\odot/\mathrm{yr}$ (adopting a clumping factor $f=0.1$), and $V_\infty$ = 1340 km/sec.  For this model we assumed a distance of 1 kpc. Changes in the distance and the mass of the central star affect the value of $\log g$ but do not appreciably modify the spectrum or the chemical abundances that we derive.  The chemical composition of the stellar wind is presented in Table \ref{tab_comp}. The He/H ratio has been reliably determined and implies that the central star is not He-rich, contrary to the results obtained by most other authors.

The chemical composition of the stellar wind may be compared with the compositions of other components in the planetary nebula system.  In particular, the iron abundance in the stellar wind is much higher than that found for the nebular gas or the hot plasma that emits in X-rays.  Therefore, it would appear that the plasma emitting diffuse X-rays in NGC 6543 must arise from heated nebular material.  The same conclusion is reached regarding the X-ray-emitting plasma in BD+30$^{\circ}$3639.  

We have also derived the chemical composition of the nebula surrounding the star based upon recombination lines (RL) and collisionally excited lines (CEL). The abundances of C, O, and Ne relative to hydrogen derived from recombination lines are from 0.4 to 0.5 dex higher than the abundances derived from collisionally excited lines.  The difference has been called the $ADF$. From five different methods involving emission lines of H, He, C, O, and Ne, we have found a mean square temperature variation $t^2= 0.028\pm 0.005$.  Supposing spatial temperature variations of this amplitude in a chemically homogeneous nebula, it is possible to reconcile the CEL and RL abundances. In this situation, we find excellent agreement between the stellar and the nebular RL abundances for He/H, C/H, and O/H.  On the other hand, the stellar N/H value is about 0.4 dex smaller than the nebular RL abundance and agrees with the nebular CEL abundance. 

This is the first paper where we make a detailed comparison between the chemical composition of the central star and of the surrounding nebula of a planetary nebula. We consider it imperative to compare the chemical composition of the central stars of planetary nebulae with those of their surrounding nebulae to advance the study of the evolution of intermediate mass stars.  This comparison is also paramount to test different hypotheses regarding the origin of a variety of observed properties, among them the large temperature variations present in many planetary nebulae and the origin of the X-ray emission. 

\acknowledgments

It is a pleasure to acknowledge Svetozar Zhekov for useful discussions. This work 
was partly supported by the CONACyT grants 60967, 46904 and 42809 and UNAM DGAPA grants IN115807, IN108406, and IN108506.



Facilities: \facility{FUSE}, \facility{HST(STIS)}, \facility{IUE}. 
\facility{SPM}



\clearpage

\clearpage


\clearpage

\begin{deluxetable}{lcccccr@{$\times$}l}
\tablecaption{Comparison of the stellar parameters obtained by different authors} 
\tablehead{
\colhead{Reference} & \colhead{$V_\infty [km/s]$} & \colhead{$\beta$} & 
\colhead{$T_{eff}\, [K]$} & 
\colhead{$L/L_\odot$} & \colhead{Distance, [pc]} & \multicolumn{2}{c}{$\dot M [M_\odot/yr]$}}
\startdata
Castor et al. (1981) & 2150 & 1.0 & 43000 & 2000 & 1170 & 9 & 10$^{-8}$ \\
Bianchi et al (1986) & 1900 & 2.0 & 80000 & 15100 & 1390 & 32 & 10$^{-8}$ \\
Perinotto at al. (1989) & 1900 & 1.5 & 60000 & 5600 & 1440 & 4 & 10$^{-8}$  \\
de Koter et al. (1996) & 1600 & 1.5 & 48000 & 5200 & - & 16 & 10$^{-8}$ \\
this work              & 1340 & 2.0 & 66750 & 1585 & 1000 & 1.86 & 10$^{-8}$ \\
\enddata
\label{tab_stellar}
\end{deluxetable}

\clearpage

\begin{deluxetable}{lccccc}
\tablecaption{Derived stellar abundances} 
\tablewidth{0pt}
\tablehead{
\colhead{Element} & \colhead{X/H} & \colhead{$12 +$ Log $N$(X)/$N$(H)} & 
\colhead{Sun\tablenotemark{a}} &
\colhead{[X/O]\tablenotemark{b}} & \colhead{Orion Nebula\tablenotemark{c}}}
\startdata
He  &  0.1       &11.00$\pm$0.04 &   10.93$\pm$0.01 &      ...     &10.988$\pm$0.003\\
C   &  1.060E-03 & 9.03$\pm$0.10 &   8.39$\pm$0.05  &+0.28$\pm$0.1 &  8.52$\pm$0.02 \\
N   &  2.292E-04 & 8.36$\pm$0.10 &   7.78$\pm$0.06  &+0.22$\pm$0.1 &  7.73$\pm$0.09 \\
O   &  1.055E-03 & 9.02$\pm$0.04 &   8.66$\pm$0.05  &      0.00    &  8.73$\pm$0.03 \\
Si  &  1.553E-04 & 8.19$\pm$0.20 &   7.51$\pm$0.04  &+0.32$\pm$0.2 &       ...      \\
Ne  &      ...   &     ...       &   7.84$\pm$0.04  &      ...     &  8.05$\pm$0.07 \\
P   &  3.390E-07 & 5.53$\pm$0.10 &   5.36$\pm$0.04  &-0.19$\pm$0.1 &       ...      \\
S   &  3.693E-05 & 7.57$\pm$0.10 &   7.14$\pm$0.05  & 0.07$\pm$0.1 &  7.22$\pm$0.04 \\
Ar  &      ...   &     ...       &   6.18$\pm$0.08  &      ...     &  6.62$\pm$0.04 \\
Fe  &  1.750E-05 & 7.24$\pm$0.10 &   7.45$\pm$0.05  &-0.57$\pm$0.1 &       ...      \\
\enddata

\tablenotetext{a}{\citet{2005ASPC..336...25A}.}
\tablenotetext{b}{[X/O] = $\log($X/O$)$-$\log($X/O$)_\odot$}
\tablenotetext{c}{\citet{2004MNRAS..355.299E}.}
\label{tab_comp}
\end{deluxetable}

\begin{deluxetable}{lccccc}
\tablecaption{Nebular collisional abundances \tablenotemark{a}
\label{tta:Col}}
\tablewidth{0pt}
\tablehead{
\colhead{Element} &
\colhead{(1)}  &
\colhead{(2)} &
\colhead{(3)} &
\colhead{(4)} &
\colhead{(5)}}
\startdata
C  &  8.50 & 8.40  &  ...  & 8.30 &  $8.40 \pm 0.10$  \\
N  &  8.50 & 8.36  &  ...  & ...  &  $8.79 \pm 0.06$  \\
Ne &  8.27 & 8.28  &  8.20 & ...  &  $8.25 \pm 0.06$  \\
S  &  7.09 & 7.11  &  7.05 & ...  &  $7.08 \pm 0.06$  \\
Ar &  6.53 & 6.62  &  ...  & ...  &  $6.57 \pm 0.06$  \\
Cl &  5.40 & ...   &  ...  & ...  &  $5.40 \pm 0.10$  \\
\enddata
\tablenotetext{a}{In units of $12 +$ Log $N$(X)/$N$(H). 1 
\citet{2004MNRAS.351..935Z}; 
2 \citet{2003A&A...406..165B}; 3 \citet{1996ASPC...99..350K}; 4 
\citet{1994A&A...282..199R} and \citet{1995RMxAA..31..131P}; 
5 adopted values.}

\end{deluxetable}

\begin{deluxetable}{lcccccc}
\tablecaption{Nebular recombination abundances \tablenotemark{a}
\label{tta:Rec}}
\tablewidth{0pt}
\tablehead{
\colhead{Element} &
\colhead{(1)}  &
\colhead{(2)} &
\colhead{(3)} &
\colhead{(4)} &
\colhead{(5)} &
\colhead{(6)}}
\startdata
He & 11.07 & 11.07 & 11.09 &...   & 11.05 & $11.05 \pm 0.01$  \\
C  &  8.90 & ...   &  8.92 & 8.87 &  ...  & $ 8.90 \pm 0.10$  \\
N  &  8.83 & ...   &  ...  & ...  &  ...  & $ 8.83 \pm 0.20$  \\
O  &  9.30 & ...   &  9.40 & ...  &  9.19 & $ 9.19 \pm 0.12$  \\
Ne &  8.67 & ...   &  ...  & ...  &  ...  & $ 8.67 \pm 0.10$  \\
\enddata
\tablenotetext{a}{In units of $12 +$ Log $N$(X)/$N$(H). 1 
\citet{2004MNRAS.351.1026W}; 
2 \citet{2003A&A...406..165B}; 3 \citet{1996ASPC...99..350K}; 4 
\citet{1994A&A...282..199R} and \citet{1995RMxAA..31..131P}; 
5 this paper; 6 adopted values.}

\end{deluxetable}

\begin{deluxetable}{lcccccc}
\tablecaption{Nebular temperatures\tablenotemark{a}
\label{tta:Tem}}
\tablewidth{0pt}
\tablehead{
\colhead{Diagnostic} &
\colhead{(1)}  &
\colhead{(2)} &
\colhead{(3)} &
\colhead{(4)} &
\colhead{(5)} &
\colhead{(6)}}
\startdata
$T$(Bac)            &  $8340 \pm 500$ & $7100 ^{+1200}_{-900}$& $6800 \pm 400$ & ... &  ... &  $6800 \pm 400$ \\
$T$(He~{\sc{ii}})   &  $6450 \pm 1000$& ...                   &...&$5730\pm 1000$& $6674 \pm 559$ & $6674 \pm 559$\\
$T$[O~{\sc{iii}}]   &  $7990 \pm 50$  & $7950 \pm 100$        &...&...  &  ...  & $7980 \pm 50$  \\
$T$[O~{\sc{ii}}]    &  $9500 \pm 500$ & ...                   &...&...&...& $9500 \pm 500$ \\
$T$([O~{\sc{iii}}],[O~{\sc{ii}}]) &...&...                    &...&...& $8100 \pm 100$& $8100 \pm 100$ \\
\enddata
\tablenotetext{a}{In K. 1 \citet{2004MNRAS.351.1026W}; 
2 \citet{1996ASPC...99..350K}; 3 \citet{2004MNRAS.351..935Z}; 4 
\citet{2005MNRAS.358..457Z}; 
5 this paper; 6 adopted values.}

\end{deluxetable}

\begin{deluxetable}{lc}
\tablecaption{$t^2$ values
\label{tta:tvalues}}
\tablewidth{0pt}
\tablehead{
\colhead{Method} &
\colhead{$t^2$}}
\startdata
$T$(Bac) and $T_e$([O~{\sc{ii}}],[O~{\sc{iii}}]) & $0.028 \pm 0.009$ \\
$T$(He~{\sc{ii}}) and $T_e$([O~{\sc{ii}}],[O~{\sc{iii}}]) & $0.035 \pm 0.014$ \\
$N$(C$^{++}$)$_{RL}$ and   $N$(C$^{+2}$)$_{CEL}$& $0.036 \pm 0.010$  \\
$N$(O$^{++}$)$_{RL}$ and   $N$(O$^{+2}$)$_{CEL}$& $0.024 \pm 0.008$  \\
$N$(Ne$^{++}$)$_{RL}$ and  $N$(Ne$^{+2}$)$_{CEL}$& $0.022 \pm 0.010$  \\
Average& $0.028 \pm 0.005$
\enddata

\end{deluxetable}

\begin{deluxetable}{lccc}
\tablecaption{Stellar and nebular abundances for NGC 6543\tablenotemark{a}
\label{tta:SteNeb}}
\tablewidth{0pt}
\tablehead{
\colhead{Element} &
\colhead{Stellar} &
\colhead{Nebular (RL)} &
\colhead{Nebular (CEL)}}
\startdata
He & 11.00  $\pm 0.04$ & $11.05 \pm 0.01$ & ...   \\
C  &  9.03  $\pm 0.10$ & $ 8.90 \pm 0.10$ & $8.40 \pm 0.10$ \\
N  &  8.36  $\pm 0.10$ & $ 8.83 \pm 0.20$ & $8.43 \pm 0.20$ \\
O  &  9.02  $\pm 0.10$ & $ 9.19 \pm 0.12$ & $8.79 \pm 0.06$ \\
S  &  7.57  $\pm 0.10$ &  ...             & $7.08 \pm 0.06$ \\
Ne &       ...         & $ 8.67 \pm$ 0.10 & $8.25 \pm 0.06$ \\
\enddata
\tablenotetext{a}{In units of $12 +$ Log $N$(X)/$N$(H).}
\end{deluxetable}

\end{document}